# Genus and spot density in the *COBE* DMR first year anisotropy maps

S. Torres[1], L. Cayón[2], E. Martínez-González[3] and J.L. Sanz[3]
[1] *Universidad de los Andes and Centro Internacional de Física, Bogotá, Colombia.*
[2] *Lawrence Berkeley Laboratory and Center for Particle Astrophysics, Berkeley, CA, USA.*
[3] *Dpto. Física Moderna, Universidad de la Cantabria and Instituto Mixto de Física de Cantabria, CSIC-U. de Cantabria, Santander (Cantabria), Spain*

**ABSTRACT**
A statistical analysis of texture on the *COBE*-DMR first year sky maps based on the genus and spot number is presented. A generalized $\chi^2$ statistic is defined in terms of "observable" quantities: the genus and spot density that would be measured by different cosmic observers. This strategy together with the use of Monte Carlo simulations of the temperature fluctuations, including all the relevant experimental parameters, represent the main difference with previous analyses. Based on the genus analysis we find a strong anticorrelation between the quadrupole amplitude $Q_{rms-PS}$ and the spectral index $n$ of the density fluctuation power spectrum at recombination of the form $Q_{rms-PS} = 22.2 \pm 1.7 - (4.7 \pm 1.3) \times n$ $\mu K$ for fixed $n$, consistent with previous works. The result obtained based on the spot density is consistent with this $Q_{rms-PS}(n)$ relation. In addition to the previous results we have determined, using Monte Carlo simulations, the minimum uncertainty due to cosmic variance for the determination of the spectral index with the genus analysis. This uncertainty is $\delta n \approx 0.2$.

## 1 INTRODUCTION

The study of texture on Cosmic Microwave Background (CMB) maps can be used to constrain scenarios of galaxy formation as an alternative technique to the temperature correlation analysis. Several statistical quantities have been proposed, such as number of spots, contour length, genus (total curvature of the iso-temperature contours), and total area of excursion above a certain level (Sazhin 1985, Bond & Efstathiou 1987, Vittorio & Juszkiewicz 1987, Coles 1988, Martínez-González and Sanz 1989, Gott et al. 1990). At large angular scales ($> 2°$) gravitational potential fluctuations at the cosmic photosphere must leave an imprint on the CMB (Sachs & Wolfe 1967), which is manifested as temperature fluctuations on the 2D sphere.

There have already been several statistical analyses of the *COBE* data. However, due to the great impact that these data have on cosmology and the difficulties of its interpretation caused by the experimental complexities and the low signal level, it is important to look at these data exhaustively. We propose and use a new statistic, based on the topological characteristics of sky maps, which is directly related with observable quantities. Analysis of the first year *COBE*-DMR maps based on the genus have recently been carried out by Torres (1994a, 1994b) and Smoot et al.

(1994). From the first paper a value of $n = 1.2 \pm 0.3$ was obtained by fitting the coherence angle of the COBE temperature maps for a fixed quadrupole-normalized amplitude of $Q_{rms-PS} = 16\mu K$, assuming a scale-free primordial spectrum of density fluctuations $P(k) \propto Q_{rms-PS}^2 k^n$. Smoot et al. (1994) allow variations of $Q$ and find a relation between $Q$ and $n$ based also on genus analysis. In Smoot et al. (1994) the covariance matrix of the genus at different temperature thresholds and for several smoothing scales is included in the $\chi^2$ minimization. The main difference of our analysis with the previous works lies in the direct comparison of the genus and spot density of the COBE data with the values obtained in each realization of a model with no additional smoothing of the data.

The previous analyses have been performed by comparison between the observed genus and the mean values deduced from the cosmological models. Mean values of model parameters are not observable quantities and therefore are not appropriate to compare with the observed data (we will see, however, that in some cases a statistic based in the mean value of a quantity can be interpreted as the mean value of the statistic based on realizations of that quantity; see section 3.1 below). In this work we use a different statistical analysis of the genus and spot density data that takes into account the ensemble of realizations that would be measured by different cosmic observers. This approach has been used by Scaramella and Vittorio (1993) with the temperature correlation function as the observable quantity. However, they do not consider all the characteristics of the *COBE*-DMR experiment such as galactic cut, pixelization, beam smearing, and smoothing, which is essential for a precise determination of the values of cosmological parameters allowed by the data.

Improvements in instrument sensitivity are very rapidly reducing experimental errors, however even in an ideal noise-less experiment the measured parametes will have the uncertainty due to cosmic variance. We have used Monte Carlo simulations to study the extent to which cosmic variance obscures the information which can be extracted from the genus analysis. A similar consideration for the rms temperature fluctuations on different angular scales results in minimal uncertainties in the determination of $n$ (White, Krauss & Silk 1993).

In section 2 we present how Monte Carlo simulations of the *COBE*-DMR maps are performed. The characteristics of the new statistical method considered in this paper are presented in section 3. The minimum range for the spectral index n, implied by cosmic variance, is obtained in section 4. In section 5 we apply the new method to the genus and spot density of the *COBE*-DMR data and show the results. An interpretation of these results in terms of the coherence angle is given in section 6. Finally, a summary of the main conclusions is given in section 7.

## 2  MONTE CARLO SIMULATIONS

Simulations that take into account instrument noise, sky coverage, galactic cut, smearing, pixelization scheme and the DMR beam characteristics were done for a set of cosmological models. Simulated maps of the cosmic signal were generated using an expansion on real harmonics for the temperature (Smoot et al. 1991) for each one of the 6144 DMR pixels:

$$T(\theta,\phi) = \sum_{l=2}^{\infty}\sum_{m=0}^{l} k\,[b_{l,m}\cos(m\phi) + b_{l,-m}\sin(m\phi)]$$
$$N_l^m W_l P_l^m(\cos\theta), \qquad (1)$$

$$N_l^m = \left[\frac{(2l+1)(l-m)!}{4\pi(l+m)!}\right]^{1/2}$$

where $k = \sqrt{2}$ for $m \neq 0$ and $k = 1$ for $m = 0$; $P_l^m(\cos\theta)$ are the Associated Legendre polynomials; the $b_{l,m}$ coefficients are real stochastic and Gaussian distributed variables with zero mean and model dependent variance $\langle b_{l,m}^2 \rangle$ given by (Abbott & Wise 1984, Bond & Efstathiou 1987)

$$\langle b_{l,m}^2 \rangle = \frac{4\pi}{5} Q_{rms-PS}^2 \frac{\Gamma(l+\frac{n-1}{2})}{\Gamma(l+\frac{5-n}{2})} \frac{\Gamma(\frac{9-n}{2})}{\Gamma(\frac{3+n}{2})}. \qquad (2)$$

The weights $W_\ell$ for DMR given by Wright et al. (1994) were used. For each realization two maps are generated by adding to the cosmic signal the noise corresponding to channels A and B of *COBE* 53 GHz. The noise is determined by instrument sensitivity and the number of observations per pixel. A small beam smearing correction is applied in order to take into account the motion of the spacecraft during the 1/2 second integration time. The two maps are added to form the $\frac{1}{2}(A+B)$ map, Gaussian smoothed ($\sigma_s = 2.9°$) and finally its genus is calculated. The algorithm used to compute the spot number density and genus is described elsewhere (Torres 1994a; 1994b).

## 3  THE NEW STATISTICAL METHOD

We define the statistic $\chi_G^2$ associated to the genus and $\chi_N^2$ associated to the number of spots as follows:

$$(\chi_G^2)^k = \sum_{i=1}^{25}\sum_{j=1}^{25}(G_i^k - G_i^{COBE})M_{ij}^{-1}(G_j^k - G_j^{COBE}). \qquad (3)$$

$G_i^k$ is the genus for the $k$-th realization at threshold level $i$, and 25 equally spaced threshold levels were used ($\nu : -3.0, ..., 3.0$); $G_i^{COBE}$ is the genus for the COBE map at threshold $i$. It is important to notice that $G_i^k$ is the genus as measured by a cosmic observer with an instrumental device like that aboard COBE and thus the statistic $(\chi_G^2)^k$ so constructed is based on observable quantities. $M_{ij}$ is the covariance matrix:

$$M_{ij} = \frac{1}{N_{realiz.}}\sum_{k=1}^{N_{realiz.}}(G_i^k - \langle G_i \rangle)(G_j^k - \langle G_j \rangle) \qquad (4)$$

The $M_{ij}$ matrix was calculated with the Monte Carlo realizations. A statistic $(\chi_N^2)^k$ related with the spot number is defined in a similar manner as $(\chi_G^2)^k$ by replacing $G_i^k$ (and similarly $G_i^{COBE}$) with $N_i^k$ ($N_i^{COBE}$), the number of spots for the realization $k$ at threshold $i$. So each model is now represented by a distribution of values $(\chi_G^2)^k$ and in order to find the model that is closest to the COBE data we have to compare distributions. The simplest way to do this is to use the mean value of the distribution $<\chi_G^2>$ (although other choises like the mode can be used they are more sensitive to the noise due to the limited number of realizations ). Notice that considering the mean value, $<\chi_G^2>$, is equivalent to a statistic defined by replacing the genus for each realization $G_i^k$ in equation (3) by the expected genus of the model, except for the constant value 25. Another possibility is to use the Kolmogorov-Smirnov statistic $K$ to compare the $\chi_G^2$ distributions. That we will do at the end of section 5 in order to check the results. For the moment we will take the simplest alternative $<\chi_G^2>$ which also requires less CPU time to compute.

## 4  IDEAL EXPERIMENT

In this section we consider an ideal experiment, that includes all the *COBE*-DMR experiment characteristics but assuming a noise-less radiometer, to elucidate the main properties of the new statistical method presented and to study the cosmic variance associated to the cosmological parameters as derived from the genus and spot number. The noise of the *COBE*-DMR radiometers is taken into account when applying the method to the real data in next section.

The no ergodic character of the CMB temperature random field on the Cosmic photosphere implies a limitation in the accuracy to determine the statistical properties of the field from a single realization. Even in the ideal case that we were able to measure the background temperature fluctuations in every part of the sphere (no galactic cut) and with negligible noise we would only be able to measure the expected values of the field within a certain error bar (cosmic variance). Uncertainties in the two point correlation function and in other higher moments due to cosmic variance have already been calculated (Scaramella and Vittorio 1990, Cayón, Martínez-González and Sanz 1991, White, Krauss and Silk 1993, Srednicki 1993, Gutiérrez de la Cruz et al. 1994).

We calculate the cosmic variance uncertainty when analysing the genus of a hypothetical sky map with the characteristics of the *COBE*-DMR experiment assuming a noise-less radiometer and no galactic contamination. In this situation the genus is independent of the amplitude of the cosmic signal and the only free parameter for the models is $n$. Thus, 7 Monte Carlo data sets were generated, one for each value of $n$ in the range 0.0-2.0. The estimated value of $n$ was obtained using the generalized $\chi^2$ method described above with each realization from model $n = 1$ taken as the input data (i.e. used as the *COBE* data in eq. 3).So, from the whole $n = 1$ data set we obtain a distribution of $n$ values. The dispersion of this distribution gives the uncertainty due to cosmic variance: $\delta n \approx 0.2$.

## 5 ANALYSIS OF THE *COBE*-DMR MAPS

### 5.1 *COBE*-DMR maps

Only data from the most sensitive radiometers were used (i.e. the DMR 53 GHz). Before any analysis was done, the released maps were processed as follows: 1. The maps were converted to thermodynamic temperature scale (a factor of 1.0742). 2. A monopole and dipole function, including the small quadrupolar component of the Doppler effect, was fitted to the maps (excluding $\pm 30°$ from the galactic plane) and subtracted. The dipole and quadrupole fit is in agreement with Smoot et al. (1992), thus providing a check for the integrity of the data and the analysis software. 3. Finally the sum $\frac{1}{2}$(A+B) and difference $\frac{1}{2}$(A-B) maps were formed and Gaussian smoothed ($\sigma_s = 2.9°$) in order to reduce noise.

### 5.2 Analysis

To find the restrictions on $Q_{rms-PS}$ and $n$ imposed by the *COBE*-DMR genus and spot density, a grid of Monte Carlo data sets were generated with $n$ in the range 0-2 (with a step of 0.2) and $Q_{rms-PS}$ between 5 and 33 $\mu$K (with a step of $2\mu$K). For each of the simulated $\frac{1}{2}$(A+B) maps, the statistics $\chi^2_G$ and $\chi^2_N$ were calculated as indicated in section 3. The number of realizations was set to 400, which proved to be a safe number after testing for the convergence of the results of the relevant quantities.

For each model (i.e. pair of values $Q_{rms-PS}, n$) one can build up, using the Monte Carlo data set, the probability $P_G(\hat{\chi}^2_G)$ of obtaining a realization with its $\chi^2_G$ smaller or equal to $\hat{\chi}^2_G$. Thus, models with a high $P_G(\hat{\chi}^2_G)$ for small $\hat{\chi}^2_G$ imply that many observers would measure a texture of their maps very similar to that of *COBE*, and the opposite for models with flatter $P_G(\hat{\chi}^2_G)$. The same can be done for the probability $P_N(\hat{\chi}^2_N)$ based on the spot number. Two examples of the distribution of $\chi^2_G$ for models $n = 0.4$, $Q_{rms-PS} = 19\mu$K and $n = 0.8$, $Q_{rms-PS} = 19\mu$K are presented in Fig. 1. For both models the histogram has a maximum near $\chi^2_G \approx 45$ and the two distributions are very similar (as we will discuss later, there is not a significant statistical difference between them).

**Figure 1.** Distributions of the $\chi^2_G$ statistic for two models: $n = 0.4$, $Q_{rms-PS} = 19$ (*solid*) and $n = 0.8$, $Q_{rms-PS} = 19$ (*dash*).

In order to establish a simple criterium to compare probabilities, and so to choose the model that best fit the data, we have used the mean $<\chi^2_G>$ ($<\chi^2_N>$) value of the realizations for each model. We then search for the minimum $<\chi^2_G>$ ($<\chi^2_N>$) in the space of model parameters ($Q_{rms-PS}, n$). The first and main result is the anticorrelation found between our estimates of $n$ and $Q_{rms-PS}$ which can be approximately given as a straight line of the form: $Q_{rms-PS} = 22.2 \pm 1.7 - (4.7 \pm 1.3) \times n$ $\mu$K for fixed $n$. An explanation of this anticorrelation in terms of the coherence angle is given in the next section. To obtain the previous relation between $Q_{rms-PS}$ and $n$, we first assign an error bar to the value of $Q_{rms-PS}$ with minimum $<\chi^2_G>$ for fixed $n$ by using the generalized $\chi^2$ method with each realization from that $Q_{rms-PS}$ taken as the input data. We then fit a straight line to the pairs ($Q_{rms-PS}, n$) with minimum $<\chi^2_G>$ considering the corresponding error bars. It is clear from this analysis that there is a wide range of values of $n$ that fit the data equally well (as can be seen in figure 1). However it seems that for the $<\chi^2_G>$ ($<\chi^2_N>$) statistic values of $n$ lower than 1 are favored. We have checked the stability of the main result with the number of realizations by performing an additional set of 400 simulations for models which differ from the minimum $<\chi^2_G>$ in 1.5 units (a number bigger than the typical error for a sampling of 400 realizations). It is verified that the ($Q_{rms-PS}, n$) relation is mantained.

The error bars for the cosmological parameters $Q_{rms-PS}, n$ were estimated by using each one of the realizations for the model $n = 0.4$, $Q_{rms-PS} = 19\mu$K (which lies in the line of degeneracy) as input data replacing COBE's data. The result is a 68% uncertainty of $\pm 0.4$ for $n$ and $^{+4}_{-6}$ for $Q_{rms-PS}$, which we consider as typical error bars for $n$ and $Q_{rms-PS}$.

Similar results were found for the number of spots $N$ but for this quantity the generalized $\chi^2$ analysis was less sensitive to variations in the model parameters.

We now test the robustness of the $<\chi^2>$ method to discriminate among the various cosmological models by comparing it with a more sophisticated method based on the Kolmogorov-Smirnov $K$ statistic. The $K$ statistic is used to compare the distribution of values $(\chi_G^2)_{COBE}^k$ obtained from equation (3) with the distribution $(\chi_G^2)_l^k$ constructed by replacing $G_i^{COBE}$ in that equation by the genus of the realization $l$ of the same model, $G_i^l$, and calculate the statistic $K_G^l$ for the two distributions. Operating in the same way with all the realizations we then obtain a distribution of $K_G^l$ values for that model and we use the mean value $<K_G>$ to compare different models. With the new $<K_G>$ minima we are able to approximately recover the relation $(Q_{rms-PS}, n)$ given above. However we do not find the trend found with $<\chi_G^2>$ favoring $n < 1$ models. Given the relatively poor signal to noise ratio of the DMR first year data there exists a range of models which are favored by the data (all the models which lie in the degeneracy law) and so a small improvement in the error bars of $<K_G>$ (associated to the limited number of realizations) to better discriminate among the models would require a large amount of CPU time (we use several ALPHA DEC 3000). However, with the 4 years COBE data set the signal to noise ratio will improve reducing the $(Q_{rms-PS}, n)$ anticorrelation and therefore making the statistic $K_G$ more suitable to discriminate among the fewer remaining models. Comparison of the $K_G$ statistic with the $<\chi_G^2>$ one shows that the former is a much better discriminator among different models.

## 6 DISCUSSION

The relation between $Q_{rms-PS}$ and $n$ found in the previous section can be interpreted in terms of the coherence angle of the temperature random field. The interpretation is based on relations between mean quantities of the field which do not take into account all the experimental restrictions and the cosmic variance. Therefore this section should be only considered as an approximate approach to the understanding of the previous results.

As a function of threshold level $\nu$, the mean value of the genus for a Gaussian random field only depends on the coherence angle of the field $\theta_c$:

$$\langle G_\nu \rangle = \left(\frac{2}{\pi}\right)^{1/2} \frac{\nu}{\theta_c^2} \exp(-\frac{\nu^2}{2}) \ , \ \theta_c = \left(-\frac{C(0)}{C''(0)}\right)^{1/2} \quad (5)$$

where the threshold level is given in terms of temperature standard deviations. The coherence angle is defined in terms of the ratio between the correlation function and its second derivative at zero lag. If it were possible to measure CMB anisotropies with ideal noise-less instruments, one could deduce from the previous expression the intrinsic coherence angle of the underlying field (if no cosmic variance and galaxy cut were present). The coherence angle of noise, as inferred from the measured genus of the $\frac{1}{2}$(A-B) map and formula (5) is $4.3° \pm 0.1°$. The coherence angle of the signal and noise 53 $\frac{1}{2}$(A+B) map derived from its genus is $4.9° \pm 0.1°$.

In the general case of a sky map including signal and noise $\theta_c$ is given by

$$\theta_c^2 = 2 \frac{\sum_l (2l+1)(C_{l,S} + C_{l,N})}{\sum_l l(l+1)(2l+1)(C_{l,S} + C_{l,N})}, \quad (6)$$

$$C_l = \langle a_l^2 \rangle \exp(-l(l+1)\sigma_{eff}^2),$$

where $\langle a_l^2 \rangle = \langle b_{l,m}^2 \rangle$ (see eq. 2), $\sigma_{eff}$ is the effective Gaussian smoothing and $C_{l,S}$ and $C_{l,N}$ are the Legendre coefficients for the signal and noise respectively (i.e. the coefficients in $C(\theta) = \frac{1}{4\pi} \sum_l (2l+1) C_l P_l(\cos\theta)$). In the extreme case of a noise-less map only the spectral form (i.e. $n$) of the model contributes to the coherence angle but not the amplitude of the signal. For instance, for the $n = 1$ model the DMR radiometers would measure an effective $\theta_c \approx 12.4°$ after beamwidth filtering ($\approx 3°$), beam smearing ($1.3°$), smoothing angle ($2.9°$) and pixel size ($1°$, corresponding to the dispersion of a Gaussian with the same area as the pixel of side $2.6°$) have been taken into account. Beam smearing is the smoothing caused by the motion of the antenna during the 1/2 second integration time per measurement. The $a_{l,N}^2$ noise coefficients can be estimated from a least squares fit of a harmonic expansion to the $\frac{1}{2}$(A-B) COBE map. A pure noise sky map would have a smaller coherence angle, $\theta_c \approx 4°.2$, considering a $\sigma_{eff}$ given by the smoothing angle, pixel size and beam smearing added in quadrature. This value agrees with the one obtained by using equation (5). Moreover, equations (2) and (6) can be solved numerically to give the theoretical $Q_{rms-PS}$ as a function of $n$ and coherence angle, $Q_{theory}(n; \theta_c)$. This law gives an anticorrelation between $Q$ and $n$ and reproduces well the empirical relation found in the previous section. Then, one can also estimate the $\theta_c$ corresponding to the 53 $\frac{1}{2}$(A+B) map by fitting the $Q_{theory}(n; \theta_c)$ to the points on the $(Q_{rms-PS}, n)$ plane with minimum $<\chi_G^2>$. This can be considered the coherence angle of the original random field and its value is $5.35° \pm 0.1°$. The anticorrelation found between $Q_{rms-PS}$ and $n$ results from the fact that models with high $n$ tend to give more weight to small angular scales, so in order to keep the same coherence angle for a given instrument noise the amplitude $Q_{rms-PS}$ of the spectrum must be decreased as $n$ increases.

Notice that by using equation (5) a different value of the coherence angle was found, as could be expected due to the uncertainty produced by the cosmic variance and the galactic cut which were not considered. The correction to this value can be calculated by Monte Carlo runs of a random field with a given $\theta_c = 5°.35$, and then comparing with the one obtained from the genus relation (eq. 5). The correction found is just the difference between the two estimated values of $\theta_c$: the value for the random field of $5°.35$ and $\theta_c$ estimated by eq. (5) for the COBE data.

Finally, we point out that a similar reasoning would apply to the results obtained from the analysis based on the number of spots $N$, simply by substituting equation (5) by the following equation:

$$\langle N_\nu \rangle = \left(\frac{2}{\pi \theta_c^2}\right) \frac{\exp(-\nu^2)}{\mathrm{erfc}(\nu/\sqrt{2})}. \quad (7)$$

## 7 CONCLUSIONS

Based on a generalized $\chi^2$ statistic which is defined in terms of the genus (or spot number) that would be measured by different cosmic observers we are able to constrain the quadrupole moment $Q_{rms-PS}$ and the spectral index $n$ of the density fluctuation power spectrum at recombination with the *COBE*-DMR first year data. We find that the two parameters should lie within a region around the straight line $Q_{rms-PS} = 22.2 \pm 1.7 - (4.7 \pm 1.3) \times n$ for fixed $n$, which corresponds to the line of constant $\theta_c = 5°.35$. This relation has been obtained with the mean value $<\chi^2>$ and confirmed with the alternative quantity $<K>$ derived from the Kolmogorov-Smirnov statistic applied to the $\chi^2$ distributions. This anticorrelation is consistent with the results of Smoot et al. (1994), using the mean genus of the models for several smoothings of the data, and with the one obtained by Seljak and Berstchinger (1993) based on an analysis of the COBE correlation function.

We have also studied the minimum uncertainty due to cosmic variance with which one can obtain the spectral index $n$ when using the genus as the statistical quantity in the comparison with the COBE data. The result is a $1\sigma$ error bar of $\delta n \approx 0.2$ (for $n \sim 1$).

## ACKNOWLEDGEMENTS

LC, EMG and JLS acknowlegde finantial support from the Human Capital and Mobility programme of the European Union, contract number ERBCHRXCT920079, and the Spanish DGICYT, project number PB92-0434-C02-02. ST was supported by the European Union under contract number CI1-CT92-0013. LC thanks the Fulbright Commission for the fellowship FU93-13924591. The COBE datasets were developed by the NASA Goddard Space Flight Center under the guidance of the COBE Science Working Group and were provided by the NSSDC.